\documentclass[pra,twocolumn,showpacs,preprintnumbers]{revtex4}
\usepackage{graphicx}
\usepackage{dcolumn}
\usepackage{bm}
\usepackage{amsmath}

\begin{document}

\title{Ground state properties of a Tonks-Girardeau Gas in a periodic
potential}
\author{Bo-Bo Wei}
\author{Shi-Jian Gu}
\author{Hai-Qing Lin}
\affiliation{Department of Physics and Institute of Theoretical Physics, The
Chinese University of Hong Kong, Hong Kong, China}

\begin{abstract}
In this paper, we investigate the ground-state properties of a
bosonic Tonks-Girardeau gas confined in a one-dimensional periodic
potential. The single-particle reduced density matrix is computed
numerically for systems up to $N=265$ bosons. Scaling analysis of
the occupation number of the lowest orbital shows that there are no
Bose-Einstein Condensation(BEC) for the periodically trapped TG gas
in both commensurate and incommensurate cases. We find that, in the
commensurate case, the scaling exponents of the occupation number of
the lowest orbital, the amplitude of the lowest orbital and the
zero-momentum peak height with the particle numbers are 0, -0.5 and
1, respectively, while in the incommensurate case, they are 0.5,
-0.5 and 1.5, respectively. These exponents are related to each
other in a universal relation.
\end{abstract}

\date{\today}
\maketitle

%71.10.Fd Lattice fermion models (Hubbard model, etc.)
%03.75.Mn Multicomponent condensates; spinor condensates
%05.70.Jk Critical point phenomena

\section{Introduction}

With the development of optical lattices and atom chip traps,
quasi-one-dimensional cold atom systems have been realized by tightly confining
the particle's motion in two directions to zero-point oscillation
\cite{Bloch,DSWeiss,DSWeiss05}. Meanwhile, by using the Feshbach resonance or
tuning the effective mass of particles moving in a periodic potential
\cite{Bloch,JLRoberts}, the inter-particle scattering length can be tuned to
almost any value desired. These progresses have led to experimental
realizations of the one-dimensional (1D) exactly solvable model that describes
an interacting Bose gas \cite{Girardeau60,Lieb1,Lieb2,Girardeau65}.

At very low temperatures and densities, a 1D Bose gas is expected to behave as
a gas of impenetrable particles known as hard-core bosons
\cite{MOlshanii98,DSPetrov00,MOlshanii01}. In particular, two recent
experiments successfully achieved the so-called Tonks-Girardeau (TG) regime and
made the TG gas a physical reality \cite{Bloch,DSWeiss}. Physically, a TG gas
is defined to be a 1D strongly correlated quantum gas consisting of bosons with
hardcore interaction. This model of a 1D Bose gas was first proposed by Tonks
in 1936 \cite{Tonks}. As a milestone development to the model, Girardeau solved
the model exactly by the famous Bose-Fermi mapping
\cite{Girardeau60,Girardeau65,Girardeau05}, where the TG gas has been mapped to
a spinless free fermion gas. Recently, it was found that the above case is just
a special case of a general mapping theorem between bosons and fermions in one
dimension, where the particles can interact with finite strength
\cite{TCheon,BEGranger}.

Since the TG gas is both theoretical exactly solvable and experimentally
accessible, there are great research interests recently in the TG gas with
different trapping potentials \cite{Girardeau2000,
Girardeau00,Girardeau01,weight02,Girardeau02,Girardeau2002,Girardeau20002,
Lapeyre02,TPapenbrock,PJForrester,AMinguzzi,DMGangardt,Wubiao,RPezer,HBuljan,
Goold, Zhangyunbo}. A review on relevant studies in this interesting subject
can be found in a recent article by Yukalov and Girardeau \cite{Girardeau05}.
In particular, Lin and Wu investigated the ground-state properties of a TG in
periodic potentials \cite{Wubiao}. They used a Monte Carlo integration
technique to compute the single-particle reduced density matrix (SPRDM) for
systems up to $N=7$ bosons. Their analysis of the ground state shows that when
the number of bosons $N$ is commensurate with the number of wells $M$ in the
periodic potential, the boson system is a Mott insulator whose energy gap is
given by the single-particle band gap of the periodic potential; however, when
$N$ is not commensurate with $M$, the system is a metal (not a superfluid). The
purpose of this work is to compute the SPRDM for large systems containing more
than 200 particles, and then study scaling relations, make quantitative
predictions for scaling exponents of the ground-state occupation numbers, and
obtain zero-momentum peak in different phases. We put particular emphasis on
examining the ground-state occupation in the two phases of the system.

The remaining of this paper is organized as follows. In Sec.~II, we introduce
the model Hamiltonian of the system and describe the single-particle
eigenstates and eigenvalues which will be used. In Sec.~III, we apply the
Bose-Fermi mapping theorem to construct the exact many-body ground state wave
function. In Sec.~IV, we devote ourselves to study the many-body properties of
the TG gas. Finally, summary and conclusions are given in Sec.~V.

\section{Model Hamiltonian and single-particle eigenstates}

\subsection{Model Hamiltonian}

We consider a gas of $N$ hardcore bosons trapped in a tight atomic waveguide.
The waveguide restricts strongly the dynamics of the gas in the transversal
directions, such that in the low temperature limit we can define our model in
the longitudinal direction only. In this direction we consider a periodic
potential such that the many-particle Hamiltonian at low density can be written
as
\begin{equation}
H={\sum_{j=1}^{N}}\left[ {-\frac{\hbar ^{2}}{2m}\frac{\partial ^{2}}{%
\partial x_{j}^{2}}+V(x_{j})}\right] ,
\end{equation}%
where $m$ is the mass of a single boson, $V(x)=V(x+d)$,  the
periodic potential with $d$ being the period. In this work, we use
the Kronig-Penney (KP) potential as a candidate for the periodic
potential, which takes the form
\begin{equation}
V(x)=\gamma^{\prime} \sum_{j=1}^{M}\delta (x-jd)
\end{equation}%
where $\gamma' $ is the strength of the $\delta$-function potential
and $M$ is the total number of periodic wells.

\subsection{Eigenstates and eigenvalues of periodic potential}

We impose the usual periodic boundary conditions and use $d$ as the unit of
distance and $\hbar ^{2}/(2md^2)$ as the unit of energy. Then we arrive at the
following dimensionless single-particle Hamiltonian,
\begin{equation}
H=-\frac{\partial ^{2}}{\partial x^{2}}+\gamma \sum_{j=1}^{M}\delta (x-jd).
\end{equation}%
where $\gamma=2md^2\gamma'/(\hbar^2 d)$. This single-particle
Hamiltonian can be solved exactly.  We review the result for later
computation. The Bloch wave functions take the form
\begin{eqnarray}
\psi _{\alpha }(x) &=&C_{\alpha }[\sin (k_{\alpha }x)+e^{-iK_{\alpha }}\sin
[k_{\alpha }(1-x)]],0\leq x\leq 1,  \notag \\
\psi _{\alpha }(x) &=&e^{iK_{\alpha }[x]}\psi _{\alpha }(x-[x]),\ 1\leq
x\leq M.
\end{eqnarray}%
The eigenenergy is%
\begin{equation}
E_{\alpha }=k_{\alpha }^{2}
\end{equation}%
where $k_{\alpha }$ satisfies the transcendental equation
\begin{eqnarray}
\cos (k_{\alpha }) &+&\gamma \frac{\sin (k_{\alpha })}{2k_{\alpha
}}=\cos (K_{\alpha }),
\end{eqnarray}%
with
\begin{eqnarray}
K_{\alpha } &=&\frac{2\pi \alpha }{M},\hspace{0.6cm}\alpha =0,\pm
1,\pm 2,\cdots \pm \frac{M-1}{2}.
\end{eqnarray}%
and the normalization constant for the Bloch wave function is given by
\begin{equation}
C_{\alpha }=\sqrt{\frac{k_{\alpha }}{M[k_{\alpha }-\frac{1}{2}\sin
(2k_{\alpha })+\cos K_{\alpha }(\sin k_{\alpha }-k_{\alpha }\cos k_{\alpha })]%
}}.
\end{equation}

\section{BOSE-FERMI MAPPING THEOREM and Many-body wave functions of the TG gas}
\label{sec:res}

For a system of $N$ identical hardcore bosons in the 1D external potential
$V(x)$, the bosons interact with each other by impenetrable pointlike
interactions, which can be more conveniently treated as a boundary condition
for the many-body wave function $\Psi _{B}(x_{1},x_{2},\ldots ,x_{N},t)$:
\begin{equation}
\Psi _{B}(x_{1},x_{2},\ldots ,x_{N},t)=0\ \text{if}\ x_{i}=x_{j},1\leq
i<j\leq N.
\end{equation}%
Then the hardcore boson gas can be considered as a free boson gas governed by
the following free Schr\"{o}dinger equation,
\begin{equation}
i \frac{\partial }{\partial t}\Psi _{B}={\sum_{j=1}^{N}}\left[
{-\frac{\partial ^{2}}{\partial x_{j}^{2}}+V(x_{j})}\right] \Psi
_{B},
\end{equation}%
where the wave function $\Psi_{B}$ satisfies the hardcore boundary condition of
Eq.~(9). Based on the observation that the hardcore boundary condition of
Eq.~(9) is automatically satisfied by a wave function of fermions due to its
antisymmetry, Girardeau \cite{Girardeau60,Girardeau65} gave the exact many-body
wave function of the hardcore boson system via the famous Bose-Fermi mapping,
which relates the wave function of hardcore bosons to that of noninteracting
spinless fermions in the same trapping potential:
\begin{equation}
\Psi _{B}(x_{1},x_{2},\ldots ,x_{N})=A\Psi _{F}(x_{1},x_{2},\ldots ,x_{N},t).
\end{equation}%
with
\begin{equation}
A=\prod_{1\leq i<j\leq N}\text{sgn}(x_{i}-x_{j}),
\end{equation}%
where sgn is sign function and $A$ is unit antisymmetric function which ensures
that $\Psi_{B}$ has proper symmetry under the exchange of two bosons. The free
fermionic wave function can be compactly written in a form of the Slater
determinant
\begin{equation}
\Psi _{F}=\frac{1}{\sqrt{N!}}\text{Det}_{m,j=1}^{N}[\psi _{m}(x_{j},t)].
\end{equation}
where $\{\psi _{m}\}, m=1,\ldots ,N$ are the single-particle
eigenstates. These eigenstates are governed by a set of uncoupled
single-particle Schr\"{o}dinger equations,
\begin{equation}
i \frac{\partial }{\partial t}\psi _{m}(x)=\left[ -\frac{\partial
^{2}}{\partial x^{2}}+V(x)\right] \psi _{m}(x).
\end{equation}%
Eqs.~(11),(13) and (14) describe how to construct the exact many-body wave
functions of a TG gas in any external potential $V(x)$. Then the wave function
of the TG gas is
\begin{equation}
\Psi _{B}=A(x_{1},x_{2},\ldots ,x_{N})\frac{1}{\sqrt{N!}}\text{Det}%
_{m,j=1}^{N}[\psi _{m}(x_{j},t)].
\end{equation}%
where $\psi _{m}(x)$ is the single-particle eigenstate in the trapping
potential $V(x)$.

In the Bose-Fermi mapping, the many-body ground state of hardcore
bosons is mapped from the many-body ground state of noninteracting
spinless fermions in the same trapping
potential\cite{Girardeau60,Girardeau65}. So to construct the ground
state wave function of a TG gas, one should choose lowest $N$
single-particle eigenstates in the slater determinant of Eq.~(15).
For the periodic potential that we are considering in this work, the
single-particle eigen functions are given by Eq.~(4), then we can
construct the exact many-body ground state wave function of the TG
gas in the KP potential according to the above procedures. The
ground-state many-body properties of the TG gas in the KP potential
can be extracted from this exact wave function.

\begin{figure}[h]
\begin{center}
\includegraphics[scale=0.85]{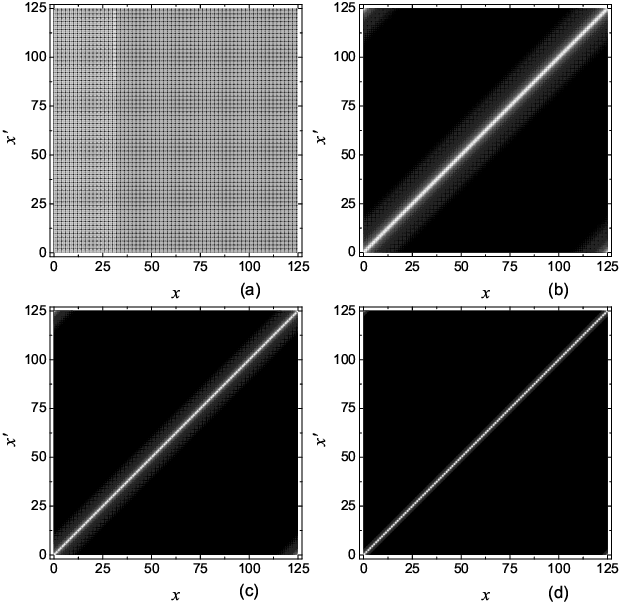}
\end{center}
\caption{Single-particle reduced density matrix,
$\protect\rho(x,x^{\prime }) $, of the TG gas in periodic potential
for different particle numbers, (a)$N=1 $, (b)$N=45$, (c)$N=85$,
(D)$N=125$ when the number of periodic wells $M=125$. All length are
in units of the period of the periodic potential $d$. }
\label{fig:epsart1}
\end{figure}

\section{Many-body properties of TG gas in periodic potentials}

In this section we investigate the ground-state properties of the 1D TG gas in
the periodic potential using the exact ground state many-body wave function of
the previous section.

\subsection{Single-particle reduced density matrix}

The many-body wave function $\Psi_B$ fully describes the state of the system.
However, its form does not transparently yield physical information related to
many important observables such as occupation numbers of natural orbitals and
the momentum distributions. The expectation value of one-body observables are
readily obtained from the SPRDM,
\begin{eqnarray}
\rho(x,x^{\prime })&=&N\int dx_2\cdots x_N \Psi_B^*(x,x_2,\ldots,x_N)  \notag
\\
&&\times \Psi_B(x^{\prime },x_2,\ldots,x_N),
\end{eqnarray}
Its diagonal elements are the position density distribution, which satisfies
the normalization condition
\begin{equation}
\int \rho(x,x)dx=N
\end{equation}

Although the exact many-body wave function of the TG gas can be written in a
compact form, the calculation of the SPRDM is a difficult task as it is very
time consuming to calculate multidimensional integrals in SPRDM for a large
system
\cite{Girardeau01,weight02,Girardeau02,Girardeau2002,Girardeau20002,Lapeyre02,
TPapenbrock,PJForrester,ALenard64,HGVaidya}. However, in the TG limit the SPRDM
can be written in a matrix product form in terms of single-particle eigenstates
\cite{RPezer}
\begin{equation}
\rho (x,x^{\prime })=\sum_{i,j=1}^{N}\psi _{i}^{\ast }(x)A_{ij}(x,x^{\prime
})\psi _{j}(x^{\prime }).
\end{equation}%
Here the $N\times N$ matrix $\mathbf{A}(x,x^{\prime })=\{A_{ij}(x,x^{\prime
})\}$ takes the form,
\begin{equation}
\mathbf{A}=(\mathbf{P}^{-1})^{T}\text{Det \textbf{P}}
\end{equation}%
where the entries of the matrix \textbf{P} are $P_{ij}(x,x^{\prime })=\delta
_{i,j}-2\int_{x}^{x^{\prime }}\psi _{i}^{\ast }(t)\psi _{j}(t)dt$, and we can
assume $x<x^{\prime }$ without loss of generality.

In addition, we observe that the SPRDM in the periodic potential we are
considering satisfies
\begin{equation}
\rho(x+1,x'+1)=\rho(x,x'),
\end{equation}
which can be easily derived from the properties of Bloch wave functions. These
formalism of the SPRDM enable us to calculate considerable large systems of the
TG gas.

The SPRDM expresses self-correlation and one can view $\rho
(x,x^{\prime })$ as the probability that, having detected the
particle at position $x$, a second measurement, immediately
following the first, will find the particle at the position
$x^{\prime }$. Classically, $\rho (x,x^{\prime })=\delta
(x-x^{\prime })$, so the off diagonal elements of SPRDM come from
purely quantum correlations of the particles. Fig.~\ref{fig:epsart1}
displays contour plots of SPRDM, $\rho (x,x^{\prime })$, for
different number of particles $N=1,45,85\ \text{and}\ 125$
respectively, where the number of wells of the periodic potential
$M=125$. We clearly see a characteristic pattern for each value of
$N$: The SPRDM are largest close to the diagonal elements which
stands for the position density distributions. The diagonal elements
of SPRDM shows oscillations due to the barrier of the periodic
potential which tends to repel the particles and push the particles
stay inside the well. The off-diagonal elements of the SPRDM relate
to off-diagonal long-range order (ODLRO) \cite{Onsager,CNYang} and
we can see that the off-diagonal elements are decreasing in contrast
to the diagonal as the number of particles $N$ increases. It means
that the repulsive interaction tends to destroy off-diagonal
coherence, so there is no ODLRO  for a system of hard core bosons in
a 1D periodic potentials in the thermodynamic limit. As $N=M$, the
SPRDM is almost diagonal and the system has diagonal long range
order while lacks ODLRO. This is because the system is a
Mott-insulator phase in the commensurate case where all the
particles are localized. In this case, we can approximate the SPRDM
as $\rho(x,x')\simeq\rho(x,x)\delta(x-x')$, which will be used in
the following sections to predict some interesting results.

\begin{figure}[h]
\begin{center}
\includegraphics[scale=0.9]{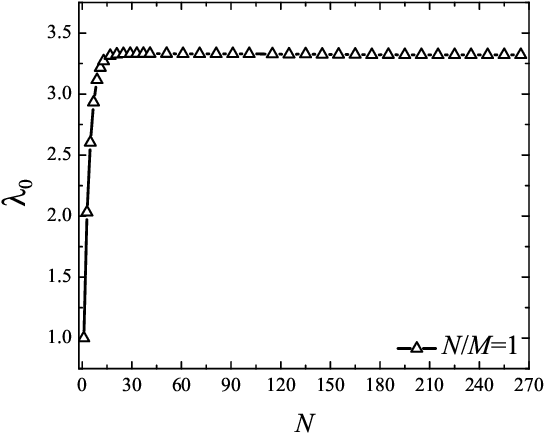}
\end{center}
\caption{Occupation numbers of the lowest orbital
$\protect\lambda_0$ as a function of particle number $N$ in the
commensurate case, $N/M=1$, when the periodic potential strength
$\protect\gamma=2$.} \label{fig:epsart2}
\end{figure}

\subsection{Occupation numbers and natural orbital}

The occupation numbers and the natural orbitals are defined from the SPRDM,
\begin{equation}
\int dx^{\prime }\rho(x,x^{\prime })\phi_i(x^{\prime
})=\lambda_i\phi_i(x), i=0,1,\ldots,
\end{equation}
where $\phi_i(x),i=0,1,2,\dots$ are the so called natural orbitals, which are
the eigenfunctions of the SPRDM and they represent effective single-particle
states. Generally, the natural orbitals are different from the single-particle
eigenstates. Unless there are some special reasons, such as translation
invariance, the single-particle eigenstates and natural orbitals are the same
plane-wave momentum eigenstates. In trapped systems, there is no simple
relation between them.  The corresponding eigenvalue $\lambda_i$ is the
occupation number of the $i$-th natural orbital. The eigenvalue of each orbital
gives the population probability of that orbital and summation of the
occupation numbers of all the orbitals satisfies $\sum_i \lambda_i=N$.  The
SPRDM is diagonal in the basis of natural orbitals, $\rho(x,x')=\sum_i
\lambda_i\phi_i^*(x)\phi_i(x')$.  The natural orbitals can be labelled in a
descending order according to their eigenvalues,
$\lambda_0>\lambda_1>\lambda_2\cdots$.

In a macroscopic interacting system, the existence of ODLRO is determined by
the behavior of $\rho (x,x^{\prime })$ as $|x-x^{\prime }|\rightarrow \infty$
\cite{Onsager,CNYang}. ODLRO is present if the largest eigenvalue of $\rho
(x,x^{\prime })$ is macroscopically non-vanishing (proportional to $N$). Which
means a macroscopic number of particles will condense to the lowest orbital. In
this case the system exhibits BEC and the corresponding eigenfunction, the
condensate orbital, plays the role of an order parameter. The fraction of
particles that are in the orbital is related to the largest eigenvalue of the
SPRDM by $f_{0}=\lambda _{0}/N$. Therefore, in analogy to the macroscopic
occupation of a single-particle eigenstate in the BEC of non-interacting Bose
gas, this orbital is sometimes referred to as the \textquotedblleft
BEC\textquotedblright state and the occupation number hence acts as a measure
of the coherence in the system.

For the TG gas in periodic potentials, there are two different phases in the
ground state \cite{Wubiao}. One is a Mott-insulator for the commensurate case
where $N/M$ is an integer. In this phase, one or more Bloch bands have been
fully occupied. The other one is a boson conductor phase for the incommensurate
case where $N/M$ is a fractional number. In this phase, the Bloch bands are
partially occupied. What we are interested in here is the scaling behavior of
the occupation number of lowest orbital in different phases. Diagonalizing the
SPRDM numerically, we can obtain the occupation numbers of the lowest orbital
for different system sizes.

\begin{figure}[h]
\begin{center}
\includegraphics[scale=0.9]{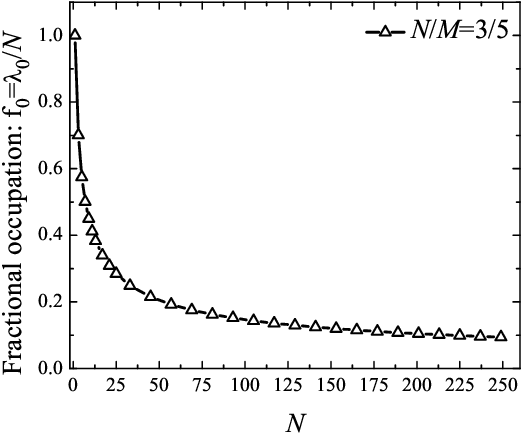}
\end{center}
\caption{Fractional occupations of the lowest orbital,
$f_0=\protect\lambda_0/N$, as a function of particle number $N$, in
the incommensurate case, $N/M=3/5$, when the periodic potential
strength $\protect\gamma=2$.} \label{fig:epsart3}
\end{figure}

\begin{figure}[h]
\begin{center}
\includegraphics[scale=0.9]{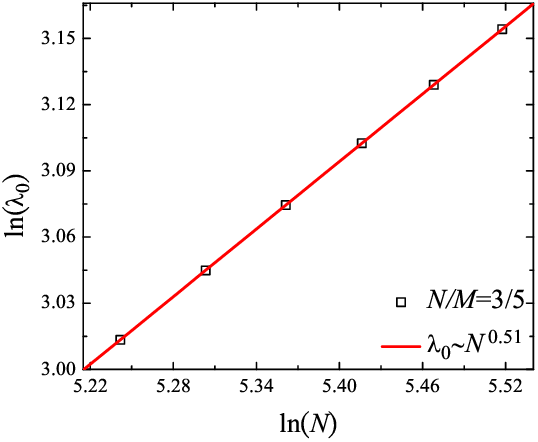}
\end{center}
\caption{(Color online) Finite size scaling analysis of the
occupation number of the lowest orbital in the incommensurate case,
$N/M=3/5$, when the periodic potential strength $\protect\gamma=2$.}
\label{fig:epsart4}
\end{figure}

We have shown the occupation number of the lowest orbital, $\lambda
_{0}$, with particle numbers $N$, in the commensurate case (the
lowest Bloch band has been fully occupied) when $\gamma =2$ in
Fig.~\ref{fig:epsart2}. From the figure, we observe that the
occupation number of the lowest orbital increases with the increase
of the system size initially and then saturate to a constant as the
particle number increases further. So the occupation number of the
lowest orbital shows $\lambda_0 \sim N^{\alpha}$ with $\alpha=0$ in
the Mott-insulating phase. The zero-exponents of the occupation
numbers of the lowest orbital can also be derived from the
approximation, $\rho(x,x^{\prime })\simeq\rho(x,x)\delta(x-x^{\prime
})$. From Eq.(21), we have
\begin{eqnarray}
\int dx^{\prime }\rho(x,x)\delta(x-x')\phi_i(x^{\prime
})&=&\lambda_i\phi_i(x),  \\
\rho(x,x)\phi_i(x)&=&\lambda_i\phi_i(x),  \\
\rho(x,x)&=&\lambda_i.
\end{eqnarray}
where $\rho(x,x)$ is the position density distribution, which are the same for
the TG gas and its fermionic counterpart,
\begin{equation}
\rho(x,x)=\sum_i |\psi_i(x)|^2.
\end{equation}
Then we have
\begin{equation}
\lambda_0=\text{Max}\{\rho(x,x)\}.
\end{equation}
The position density distributions satisfy $\int \rho(x,x)dx=N$. Meanwhile, the
position density distributions are periodic with period of 1 in the periodic
potential, which can be clearly observed from Fig.~\ref{fig:epsart1}. So the
position density distributions are normalized inside each well, which is
independent of the system size in the commensurate case. Then the maximum of
$\rho(x,x)$ is also independent of the system size,
$\text{Max}\{\rho(x,x)\}\sim N^0$. Thus we have
\begin{equation}
\lambda_0\sim N^0.
\end{equation}
So this confirms our numerical result, $\lambda_0 \sim N^0$, which
is shown in Fig.~\ref{fig:epsart2}, in the Mott-insulating phase.
Thus we can infer that there is no BEC for the periodically trapped
TG gas in the commensurate case.

The fractional occupation of the lowest orbital, $f_{0}=\lambda
_{0}/N$, as a function of particle number $N$, in the incommensurate
case, $N/M=3/5$, and $\gamma =2$ is displayed in
Fig.~\ref{fig:epsart3}. We see that the fractional occupation of the
lowest orbital in the incommensurate case decreases with the
increases of the particle number and will vanish in the
thermodynamic limit. Doing finite size scaling of the occupation
number of the lowest orbital in Fig.~\ref{fig:epsart4}, from which
we find that the occupation numbers of the lowest orbital $\lambda
_{0}$ shows power law dependence on the particle number $N$ of the
system in the incommensurate case, i.e $\lambda _{0}\sim N^{\alpha}$
with $\alpha=0.51$ for the number of particles up to $N=249$. In
Table I, we have made a list of the best converged exponents and the
number of particles involved. Simple scaling arguments presented
below show that the exponents will converge to 0.5 in the
thermodynamic limit. This behavior of periodically trapped hard-core
bosons in the incommensurate case is similar to the uniform system
of hard-core bosons \cite{PJForrester,ALenard64} and harmonically
trapped hardcore bosons \cite{TPapenbrock,PJForrester}. So there is
also no BEC behavior for periodically trapped hardcore bosons in the
incommensurate case.

\begin{figure}[h]
\begin{center}
\includegraphics[scale=0.9]{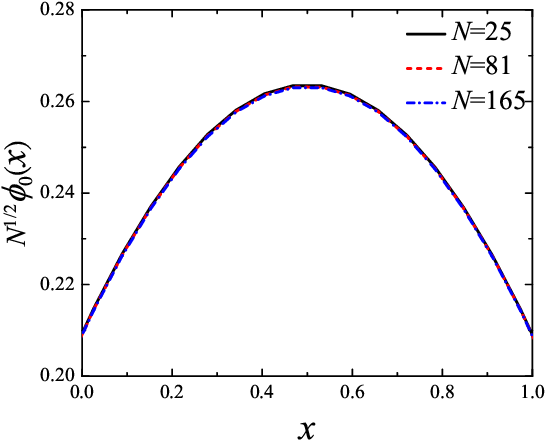}
\end{center}
\caption{(Color online)  Scaled lowest natural orbital
$\sqrt{N}\phi_0(x)$  for $N=25$(full line), $N=81$ (dashed line),
and $N=165$ (dashed-dotted line) bosons in the incommensurate case,
$N/M=3/5$, when the periodic potential strength $\protect\gamma=2$.
The amplitude of the lowest natural orbitals are in units of
$1/\sqrt{d}$ and the length is in units of the period of the
periodic potential $d$. } \label{fig:epsart5}
\end{figure}

\begin{figure}[h]
\begin{center}
\includegraphics[scale=0.9]{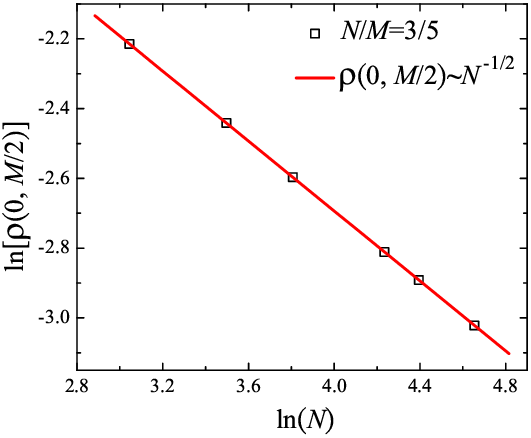}
\end{center}
\caption{(Color online) Off-diagonal element $\rho(0,M/2)$ as a
function of particle number $N$ in the incommensurate case,
$N/M=3/5$, when periodic potential strength $\protect\gamma=2$. The
off-diagonal element $\rho(0,M/2)$ is in units of the inverse of the
period of the periodic potential $1/d$. } \label{fig:epsart6}
\end{figure}

We display the scaled lowest natural orbital $\sqrt{N}\phi_0(x)$ for
$N=25,81,165$ in the incommensurate case in Fig.~\ref{fig:epsart5}.
The lowest natural orbital in the incommensurate case are periodic
with period of 1, so we show the natural orbitals in one period
only. We observe that the scaled lowest natural orbitals are
$N$-independent function that depends only on the variable $x$ as
$N$ increases. We find that the amplitude of the lowest natural
orbital in the incommensurate case scales as $\phi_0(x)\propto
N^{\beta}$ with $\beta=-0.5$.  This scaling behavior is expected as
the natural orbitals are normalized and extend from $0$ to $M$(in
the order of $N$). Similarly, the lowest natural orbital in the
commensurate case is also periodic with period of 1 due to the
presence of periodic potential. The natural orbitals are all
normalized, $\int|\phi_0(x)|^2dx=1$, then the amplitude of the
lowest orbital also scales as $\phi_0(x)\propto N^{\beta}$ with
$\beta=-0.5$ in the commensurate case.

The scaling behavior of diagonal elements of SPRDM is $N^0$, as
$\int \rho(x,x)dx=N $ and $x$ extends over an interval in the order
of $N$. We know the off-diagonal elements qualitatively decrease as
$N$ increases from Fig.~\ref{fig:epsart1}, to see this more clearly,
we do finite size scaling of the off-diagonal elements,
$\rho(0,M/2)$ , of SPRDM in Fig.~\ref{fig:epsart6}. It is clear from
the figure that, with the increases of the particle number $N$, the
interaction in the system gets stronger, then the off-diagonal
element decreases. This indicates that the interaction suppresses
the off-diagonal terms. Secondly, we find that off-diagonal element
scales as $\rho(0,M/2)\propto N^{-1/2}$ with the particle number
$N$.

These findings are particularly interesting, because they allow us to predict
the scaling behavior of the occupation number of the lowest orbital $\lambda_0$
in the incommensurate case. From Eq.(21), we have
\begin{eqnarray}
\lambda_0 \phi_0(0)=\int \rho(0,x')\phi_0(x')dx' &\sim&
N^{-0.5}N^{-0.5}\times N \nonumber\\
&\sim&  N^0
\end{eqnarray}
where $\phi_0(0)$ is non-zero as the finite strength of the periodic potential
we are considering, which also scales as $\phi_0(0)\sim N^{-0.5}$. Therefore,
we have obtained the scaling behavior of the occupation number of the lowest
orbital in the incommensurate case,
\begin{eqnarray}
\lambda_0 \sim \sqrt{N}.
\end{eqnarray}

\begin{table}[ht]
\begin{tabular} {|c|c|} \hline
Number of particles $N$  & Scaling expoents $\alpha$ ($\lambda_0\sim
N^{\alpha}$)
\\ \hline  25& 0.56
\\ \hline 45 &  0.54
\\ \hline 105 & 0.52
\\ \hline  249  & 0.51
\\ \hline  $\infty$  & 0.5\\
\hline
\end{tabular}
\caption{\label{tab:table1} Comparison of best converged scaling exponents of
the occupation number of the lowest natural orbital in the incommensurate case
and particle number involved.}
\end{table}

\begin{figure}[h]
\begin{center}
\includegraphics[scale=0.9]{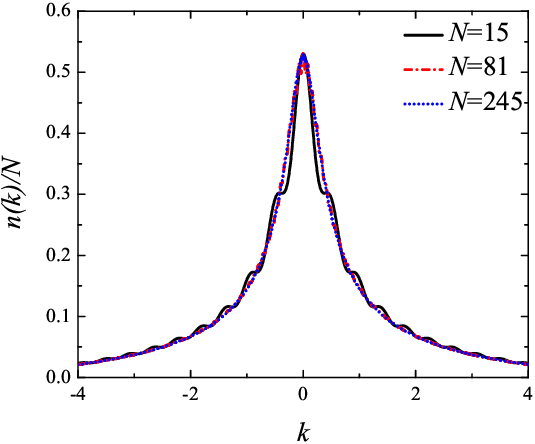}
\end{center}
\caption{(Color online) Normalized momentum density distributions
for $N=15$ (solid line), $N=81$(dashed-dotted line) and $N=245$
(dotted line) bosons in the commensurate case, $N/M=1$, when the
periodic potential strength $\protect\gamma=2$. The momentum $k$ is
given in units of the inverse of the period of the periodic
potential, $1/d$,  and the momentum density is in units of the
period of the periodic potential $d$. } \label{fig:epsart7}
\end{figure}

\begin{figure}[h]
\begin{center}
\includegraphics[scale=0.9]{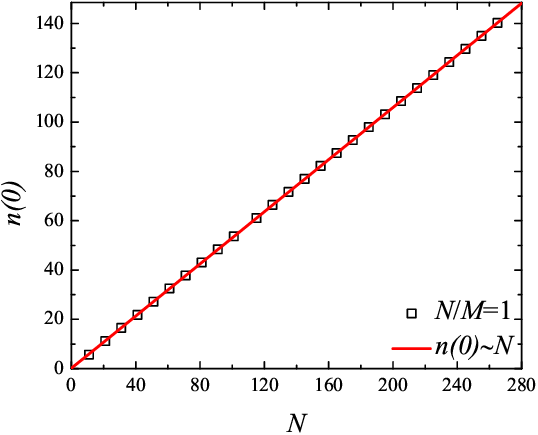}
\end{center}
\caption{(Color online) Zero-momentum peak, $n(0)$, as a function of
particle number $N$ in the commensurate case, $N/M=1$, when the
periodic potential strength $\protect\gamma=2$. $n(0)$ is in units
of the period of the periodic potential $d$. } \label{fig:epsart8}
\end{figure}

\subsection{Momentum density distribution}

Although the position density profiles and energy spectrum are exactly the same
between the hardcore boson gas and its corresponding spinless free fermion gas
due to Bose-Fermi mapping theorem, the momentum density distributions differ
considerably from each other \cite{Wubiao,Zhangyunbo}. The momentum density
distributions can be obtained from the SPRDM as
\begin{equation}
n(k)=(2\pi)^{-1}\int dx dx^{\prime }\rho(x,x^{\prime})
e^{-ik(x-x^{\prime })}.
\end{equation}
Obviously, the momentum density distributions satisfy
\begin{equation}
\int n(k)dk=N.
\end{equation}

The normalized momentum density distributions of periodically
trapped TG gas in the commensurate case are shown in
Fig.~\ref{fig:epsart7} for various particle numbers. We see that the
normalized momentum density distributions of the TG gas has a
bosonic structure, which has peaks at $k=0$ and the profiles are
narrow. While its fermionic counterpart, the spinless free fermion
gas, has a broad Fermi momentum distributions. Although a TG gas has
the same energy spectrum and position profiles with its fermionic
counterpart, their momentum distributions are very different. In
addition, the profiles of normalized momentum distributions are
nearly unchanged with the increase of the system sizes in the
commensurate case. This is because the periodically trapped TG gas
in the commensurate case is a Mott-insulating phase, so particles
are localized in the real space. In the language of matter waves,
the wave packages of the particles have no overlapping. With the
increase of system size, the particles do not affect each other in
any essential way. The oscillations in the momentum density
distributions come from the finite size effect, it will vanish in
the thermodynamic limit. What is more interesting here is that we
find the momentum density distributions have same zero momentum
peaks after rescaling. Then we plot $n(0)$ as a function of particle
number $N$ in Fig.~\ref{fig:epsart8}. It clearly shows that
$n(0)\propto N^{\gamma}$ with $\gamma=1$ in the commensurate case.
Similar behavior has been found for the TG gas in the harmonic trap
\cite{TPapenbrock}. In the commensurate case, which is the
Mott-insulating phase, we can approximate the SPRDM as
$\rho(x,x')\simeq\rho(x,x)\delta(x-x')$, then the zero-momentum
density can be obtained as
\begin{eqnarray}
n(0)&=&(2\pi)^{-1}\int dx dx^{\prime }\rho(x,x^{\prime}) \\
    &\simeq&(2\pi)^{-1}\int dx dx^{\prime }\rho(x,x)\delta(x-x^{\prime}) \\
    &\simeq&(2\pi)^{-1}\int dx \rho(x,x)\\
    &\simeq&(2\pi)^{-1}N
\end{eqnarray}
which confirms our numerical result shown in Fig.~\ref{fig:epsart8}.

\begin{figure}[h]
\begin{center}
\includegraphics[scale=0.9]{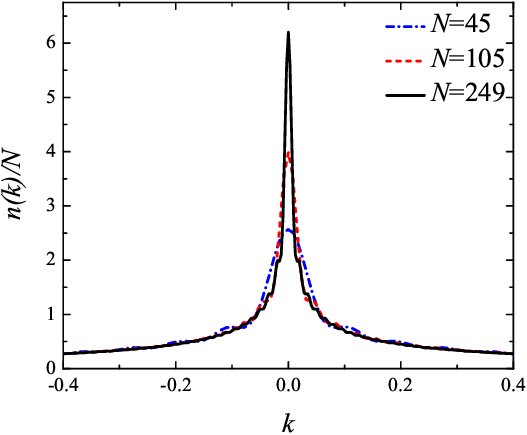}
\end{center}
\caption{(Color online) Normalized momentum density distributions
for $N=45$ (dash-dotted line), $N=105$ (dashed line) and $N=249$
(solid line) bosons in the incommensurate case, $N/M=3/5$, when the
periodic potential strength $\protect \gamma=2$. The momentum $k$ is
given in units of the inverse of the period of the periodic
potential, $1/d$,  and the momentum density is in units of the
period of the periodic potential $d$.} \label{fig:epsart9}
\end{figure}

\begin{figure}[h]
\begin{center}
\includegraphics[scale=0.9]{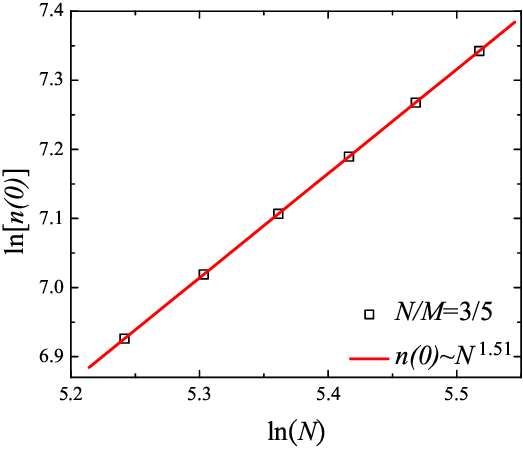}
\end{center}
\caption{(Color online) Zero-Momentum peak, $\ln[n(0)]$, as a
function of particle number, $\ln(N)$, in the incommensurate case,
$N/M=3/5$, when the periodic potential strength $\protect \gamma=2$.
$n(0)$ is in units of the period of the periodic potential $d$. }
\label{fig:epsart10}
\end{figure}

The normalized momentum density distributions of periodically
trapped TG gas in the incommensurate case are shown in
Fig.~\ref{fig:epsart9} for various particle numbers. We see that, in
the incommensurate case, the profiles of the normalized momentum
density distributions become narrower with the increase of the
system sizes, which is different from that in the commensurate case
we discussed above. This is because in the incommensurate case, the
system is a boson conductor, so the spatial distribution become more
uniform as the system size increases due to many body repulsions of
the particles. The small oscillations in the momentum profiles come
from finite size effects and it will vanish in the thermodynamic
limit. We do finite size scaling of zero-momentum peak with the
particle number in the incommensurate case in
Fig.~\ref{fig:epsart10}, and we found, up to particle number
$N=249$, the zero-momentum peak $n(0)\propto N^{\gamma}$ with
$\gamma=1.51$. A universal relation among the scaling exponents of
the ground-state occupation, the amplitude of the lowest natural
orbital, and the zero-momentum peak height presented below shows
that the exponent of the zero-momentum peak will converge to 1.5 for
the incommensurate case in the thermodynamic limit.

To find a universal relation among various exponents presented above, we follow
similar arguments of reference \cite{TPapenbrock}. The zero momentum peak is
calculated as
\begin{eqnarray}
n(0)&=& (2\pi)^{-1}\int dx dx^{\prime }\rho(x,x^{\prime})
    \\
&=& (2\pi)^{-1}\int dx dx^{\prime }\sum_j \lambda_j
\phi_j(x)\phi_j(x') \\ &=& (2\pi)^{-1}\int dx
dx^{\prime}\lambda_0 \phi_0(x)\phi_0(x^{\prime})\\
 & &+(2\pi)^{-1}\int dx dx^{\prime }\sum_{j>0} \lambda_j
\phi_j(x)\phi_j(x')\\
    &\propto& \lambda_0\int dx dx^{\prime
    }\phi_0(x)\phi_0(x^{\prime}).
\end{eqnarray}
Here we first express the SPRDM in the basis of natural orbitals where it is of
a diagonal form. Then we discarded the contributions from higher orbitals since
the occupations of higher orbitals are small. Meanwhile, the lowest orbital has
even parity (no node) while the higher orbitals have node, so the integral of
higher orbitals leads to cancellations.

If we define $n(0)\sim N^{\gamma}$, $\lambda_0\sim N^{\alpha}$ and $\phi_0(x)
\sim N^{\beta}$, then we have a universal relation among these exponents,
\begin{equation}
\gamma=\alpha-2\beta.
\end{equation}
Similar relation has been found in harmonically trapped hardcore
boson gas \cite{TPapenbrock}. However, we would like to stress that
the scaling relation we found in Eq.(41) is more general, as it is
valid for hardcore boson gas in various trapping potentials.

\begin{table}[h]
\begin{tabular} {|c|c|c|c|} \hline
{Scaling Exponents}  & $\alpha$ & $ \beta$ & $\gamma$
\\[1.0ex] \hline Ideal Bose gas& 1 & -0.5 & 2
\\[1.0ex] \hline Uniform TG gas\cite{ALenard64,PJForrester} & 0.5 & -0.5 &
1.5
\\[1.0ex] \hline Harmonically trapped TG gas\cite{TPapenbrock,PJForrester} & 0.5 & -0.25 & 1
\\[1.0ex] \hline Periodically trapped TG gas(Mott phase) & 0   & -0.5 & 1
\\[1.0ex] \hline Periodically trapped TG gas(Superfluid phase) & 0.5 &  -0.5
&1.5\\[1.0ex]
\hline
\end{tabular}
\caption{\label{tab:table1}Comparison of scaling exponents of the occupation
number of the lowest orbital, the amplitude of the lowest orbital and the
zero-momentum peak height for the TG gas in different trapping potentials.}
\end{table}

We summarize the various scaling exponents for the hardcore bosons in different
trapping potential in Table II. From Table II, we see that the scaling
exponents of TG gas in different trapping potentials are quite different but
these exponents are related to each other in a universal relation Eq.~(41).

One may be confused why $n(0)\propto N^2$ rather than $n(0)\propto
N$, for the ideal Bose gas, this comes from the difference between
continuous case and discrete case,
\begin{equation}
\int n(k)dk=N,
\end{equation}
while
\begin{equation}
\sum_k n_k=\sum_k \langle a_k^\dagger a_k\rangle=N.
\end{equation}
So in the thermodynamic limit, where $N\rightarrow \infty,
L\rightarrow \infty$ and $N/L\rightarrow \text{const}$, we have
\begin{equation}
n(k)=\frac{L}{2\pi}n_k\propto N n_k=N \langle a_k^\dagger
a_k\rangle,
\end{equation}
Therefore for the ideal gas,
\begin{equation}
n(0)\propto N\langle a_0^\dagger a_0\rangle \propto N^2.
\end{equation}

\section{Summary and Conclusions}
\label{sec:sum}

We have investigated the ground-state properties of a 1D TG gas trapped in a
periodic potential. Based on the exact many-body wave function, obtained from
Bose-Fermi mapping theorem, we have calculated the SPRDM numerically for
systems containing more than 200 particles by employing the technique that the
SPRDM can be expressed in a matrix product form in the TG gas limit. We have
observed that, with the particle number increases, the SPRDM is almost diagonal
which lacks ODLRO. The scaling analysis of the occupation numbers of the lowest
orbital shows that there are no BEC for the periodically trapped TG gas both in
the commensurate case and in the incommensurate case. We found that, in the
commensurate case, the scaling exponents of the occupation numbers of the
lowest orbital, the amplitude of the lowest orbital and the zero-momentum peak
height scale with the particle numbers are 0, -0.5 and 1 respectively, while
are 0.5, -0.5 and 1.5 respectively in the incommensurate case. These exponents
are related to each other by a simple relation.

\begin{acknowledgements}
We thank Prof.~B.~Wu and Y. F. Lo for valuable discussions. We are grateful to
D. P. Zhang for his critical reading of our manuscript. This work is supported
by RGC Grant CUHK 402107 and CUHK 401108.
\end{acknowledgements}

\end{document}